\documentclass[sigconf]{acmart}
\AtBeginDocument{%
  \providecommand\BibTeX{{%
    \normalfont B\kern-0.5em{\scshape i\kern-0.25em b}\kern-0.8em\TeX}}}

\copyrightyear{2024}
\acmYear{2024}
\setcopyright{acmlicensed}\acmConference[ICSE '24]{2024 IEEE/ACM 46th International Conference on Software Engineering}{April 14--20, 2024}{Lisbon, Portugal}
\acmBooktitle{2024 IEEE/ACM 46th International Conference on Software Engineering (ICSE '24), April 14--20, 2024, Lisbon, Portugal}
\acmDOI{10.1145/3597503.3639154}
\acmISBN{979-8-4007-0217-4/24/04}

\usepackage{listings}

\usepackage{tabularx}
\usepackage{enumitem}

\usepackage{subcaption}
\usepackage[automake,toc,abbreviations]{glossaries-extra}
\usepackage{marginnote}

\definecolor{commentcolor}{rgb}{0.13,0.55,0.13}
\definecolor{cellcolor}{RGB}{220,220,220}
\definecolor{b_color}{RGB}{247, 176, 176}
\lstdefinestyle{mystyle}{
    language=C,
    basicstyle=\ttfamily\footnotesize,
    keywordstyle=\color{black}\ttfamily,
    stringstyle=\color{black}\ttfamily,
    commentstyle=\color{black},
    morecomment=[l][\color{magenta}]{\#},
    showstringspaces=false,
    numbers=left,
    numberstyle=\tiny,
    frame=single,
    breaklines=true
}

\newglossaryentry{computer}
{
name=computer,
description={A programmable machine that receives input data,
               stores and manipulates the data, and provides
               formatted output}
}

\newglossary*{nomenclature}{Nomenclature}
\newglossaryentry{dingledorf}
{
type=nomenclature,
name=dingledorf,
description={A person of supposed average intelligence who makes incredibly brainless misjudgments}
}

\newabbreviation[longplural={Code Generation Tools}]{cgt}{CGT}{Code Generation Tool}
\newabbreviation[longplural={Common Weakness Enumerations}]{cwe}{CWE}{Common Weakness Enumeration}
\newabbreviation{cve}{CVE}{Common Vulnerabilities and Exposures}
\newabbreviation[longplural={Large Language Models}]{llm}{LLM}{Large Language Model}
\newabbreviation[longplural={Recurrent Neural Networks}]{rnn}{RNN}{Recurrent Neural Network}
\newabbreviation{nlp}{NLP}{Natural Language Processing}
\newabbreviation[longplural={Long Short-Term Memories}]{lstm}{LSTM}{Long Short-Term Memory}

\makeglossaries

\begin{document}

\title{A User-centered Security Evaluation of Copilot}

\author{Owura Asare}
\affiliation{%
  \institution{University of Waterloo}
  \city{Waterloo}
  \country{Canada}}
\email{oasare@uwaterloo.ca}

\author{N. Asokan}
\affiliation{%
  \institution{University of Waterloo}
  \city{Waterloo}
  \country{Canada}}
\email{asokan@acm.org}

\author{Meiyappan Nagappan}
\affiliation{%
  \institution{University of Waterloo}
  \city{Waterloo}
  \country{Canada}}
\email{mei.nagappan@uwaterloo.ca}

\renewcommand{\shortauthors}{Asare et al.}

\begin{abstract}
  Code generation tools driven by artificial intelligence have recently become more popular due to advancements in deep learning and natural language processing that have increased their capabilities. The proliferation of these tools may be a double-edged sword because while they can increase developer productivity by making it easier to write code, research has shown that they can also generate insecure code. In this paper, we perform a user-centered evaluation GitHub's Copilot to better understand its strengths and weaknesses with respect to code security. We conduct a user study where participants solve programming problems (with and without Copilot assistance) that have potentially vulnerable solutions. The main goal of the user study is to determine how the use of Copilot affects participants' security performance. In our set of participants (n=25), we find that access to Copilot accompanies a more secure solution when tackling harder problems. For the easier problem, we observe no effect of Copilot access on the security of solutions. We also observe no disproportionate impact of Copilot use on particular kinds of vulnerabilities. Our results indicate that there are potential security benefits to using Copilot, but more research is warranted on the effects of the use of code generation tools on technically complex problems with security requirements.
\end{abstract}



\maketitle

\section{Introduction}\label{sec:intro}

\glspl{cgt} have recently become more popular due to their ability to make developers more productive during the software development process. By \gls{cgt}, we refer to \glspl{llm} and their fine-tuned descendants that are used to generate code. These tools have improved as a result of progress in deep learning and natural language processing that have made it possible to train increasingly more capable large language models in an efficient manner. 

\glspl{cgt} have the ability to become large scale producers of insecure code if left to grow unchecked.
This is because \glspl{cgt} are trained on code sourced from repositories that lack security guarantees, making it probable that they are trained on code segments with security vulnerabilities.
This in turn leads to the possible production of insecure code when these \glspl{cgt} are used by developers.
Empirical evidence by Pearce et al.~\cite{pearce_asleep_2022} substantiates this claim, revealing that GitHub's Copilot\cite{github_inc_github_2021}, a widely-used \gls{cgt} based on the Codex \gls{llm}\cite{chen_evaluating_2021}, produces insecure code approximately 40\% of the time.

To mitigate the possible adverse effects of \glspl{cgt}, it is important that we gain a deeper understanding of their impacts on security.
In this paper, we present our work on a user-centered security evaluation of GitHub's Copilot that aims to provide a better understanding of how \glspl{cgt} affect code security. This study serves as a non-exact replication of other studies that have also conducted investigations of the security of \glspl{cgt} through user studies, specifically the work by Sandoval et al.\cite{sandoval_lost_2023} and the work by Perry et al.\cite{perry_users_2022}. 
The work by Pearce et al. \cite{pearce_asleep_2022} demonstrated that Copilot generates vulnerabilities when used as a standalone code completion tool (i.e. the code generated by Copilot was unedited). Our goal in this study is to find out whether the use of Copilot and the ability to edit its suggestions results in code with more or fewer vulnerabilities.
A study of this nature is warranted because while \glspl{cgt} become more popular among developers, their security capabilities remain under-explored and the differing conclusions of the previously mentioned studies suggest a lack of a consensus in the literature on the security effects of using them. Non-exact replication studies like ours play an important role in empirical science. It helps in generalizing results to a broader extent and contributes to the cumulative nature of scientific knowledge. In our study, we investigate how Copilot (when used as an assistant) affects users' security performance by designing and conducting a user study where participants solve programming problems with and without the assistance of Copilot.
We observe that participants in our sample generally have a better security performance when access to Copilot is granted for difficult problems, and less so for relatively simpler problems. We also observe a more uniform performance across the different types of vulnerabilities when Copilot is in use. By this we mean that the presence of Copilot does not seem to disproportionately influence the presence or absence of any type of vulnerability.
Overall, our results suggest that:

\begin{enumerate}
    \item the security benefits of using Copilot are more noticeable when it is used for more complex problems (Section \ref{sec:rq1})
    \item the use of Copilot does not necessarily increase or reduce the chances of any particular vulnerability (Section \ref{sec:rq2})
\end{enumerate}


We provide access to all our study material at \url{https://github.com/ppdb1123/copilot-user-study-supp}
\section{Background}\label{sec:background}

\subsection{Language Models and Code Generation}
Language models are generally defined as probability distributions over sequences of words. Language models model language by probabilistically predicting/generating the next word in a given sequence. They are able to do this by leveraging a set of parameters that are obtained after training on significant amounts of data. Over time, language models have evolved, from non-neural models (N-grams), to neural, recurrence-based models (RNNs)~\cite{bengio_neural_2000}, to attention-based models (Transformers)~\cite{vaswani_attention_2017}. 

The evolution and rise in popularity of language models have led to their applications to several tasks across many domains. One such domain is software engineering where language models have been put to the task of code generation. \glspl{cgt}, available either through integrated development environments (IDEs) or as extensions to text editors, are already widely used by developers~\cite{dohmke_github_2022} and they continue to evolve in complexity. GitHub’s Copilot~\cite{github_inc_github_2021} is an example of an evolved \gls{cgt}. Copilot is generally described as an AI pair programmer trained on billions of lines of public code. Currently available as an extension for the VSCode text editor, Copilot takes into account the surrounding context of a program and generates possible code completions for the developer. IntelliCode~\cite{svyatkovskiy_intellicode_2020} is another example of a \gls{cgt} that generates recommendations based on thousands of open-source projects on GitHub.

Most current high performing models use the Transformer model which was initially introduced with two components: an encoder and a decoder. There are, however, high performing models that either only use the encoder~\cite{devlin_bert_2019} or the decoder~\cite{brown_language_2020}. Copilot is based on OpenAI's Codex~\cite{chen_evaluating_2021}, which is itself a fine-tuned version of GPT-3~\cite{brown_language_2020}.


\section{Research Overview}\label{sec:overview}

\subsection{Motivation}

\glspl{cgt} are designed to assist programmers during the code writing phase of the software development process. In this paper, we are interested in how Copilot affects the security of code written by the human participants in our user study. While there is existing research on evaluating LLM-based code assistants~\cite{sandoval_lost_2023, perry_users_2022}, ours is the first to focus on Copilot which is the more mature and fine-tuned \gls{cgt}. Copilot is also more popular and more accessible, which makes our findings more likely to be applicable to developers' experience in non-experimental settings.

\subsection{Research Questions}

\begin{enumerate}
    \item Does Copilot use correlate with participants writing more secure code?
    \item Are there vulnerability types that Copilot is more susceptible to or more resilient against?
\end{enumerate}

\section{Method}\label{sec:methodology}

\begin{figure*}
    \includegraphics[scale = 0.10]{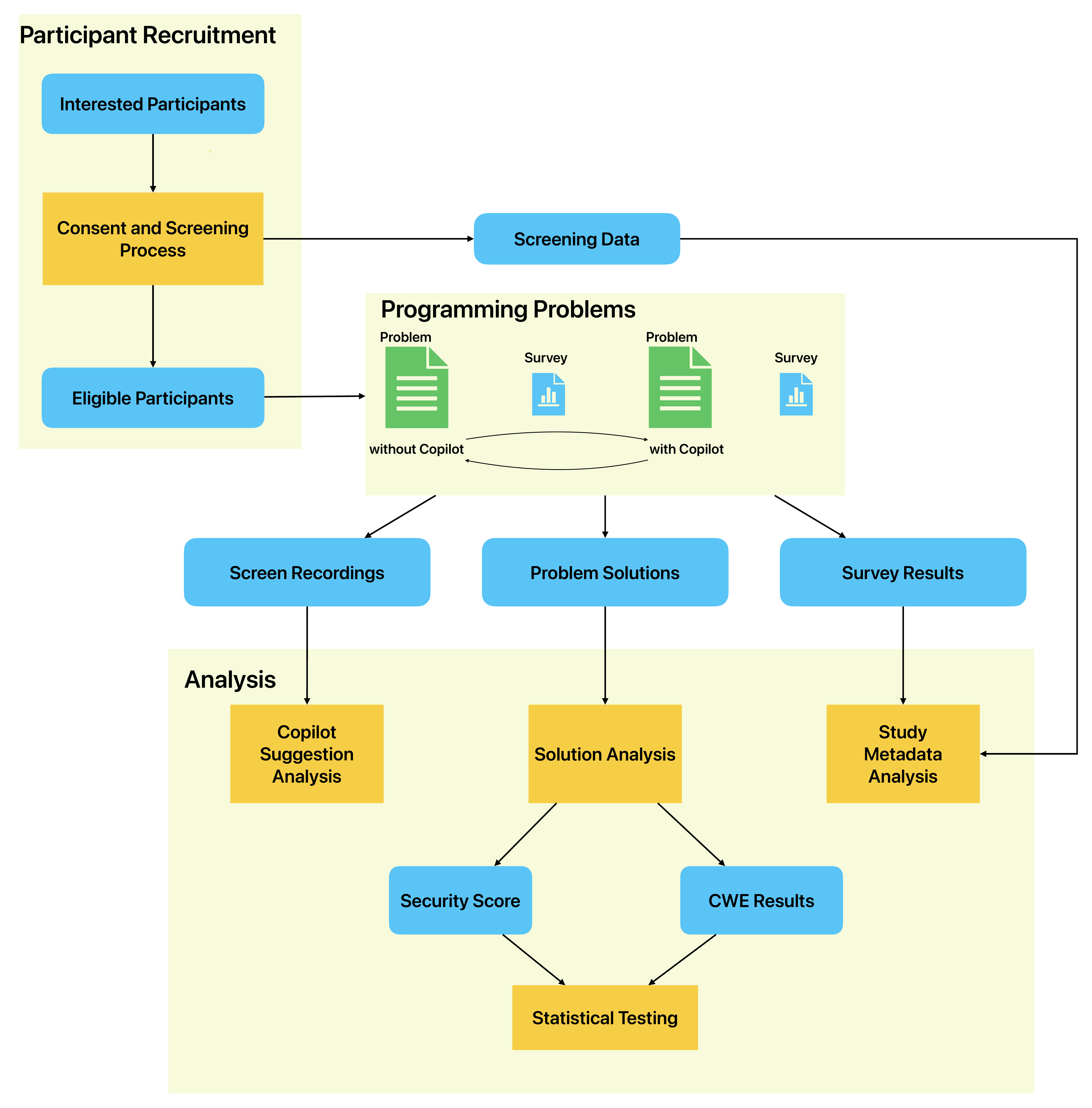}
    \caption{An overview of the user study, highlighting the key steps from recruiting participants to analyzing results.}
    \label{fig:user-flowchart}
\end{figure*}

Here we discuss our approach for the study. Figure \ref{fig:user-flowchart} summarizes our method.

\subsection{Participants: Recruitment and Screening}

Participants for this study were recruited online via mailing lists. While our main source of participants was the University of Waterloo computer science graduate student mailing list, we also extended invitations to industry professionals and potentially qualified undergraduate students. Participants who expressed interest in the study were asked to fill consent and screening forms which we used to determine their eligibility for the study. Selection criteria for this study was based on age (over 18 years), programming experience (at least one year of programming experience in C/C++), access to Copilot, and employment history (no affiliation with the development of Copilot, GitHub, or OpenAI). Participants who met our selection criteria were allowed to schedule a two hour online study session for the experiment to be conducted.

Like prior studies in this area, participants were not screened based on security experience because the goal of this study was to investigate the impact of Copilot use on code security among ordinary developers, regardless of security experience. We do however consider the idea of exploring the effects of participant security experience on the security of code generated with CGT assistance an interesting avenue for future research.

Overall, 33 people expressed interest in taking part in the study.
8 of them either did not complete the consent and screening process or did not select a time for the problem solving session.
25 out of the 33 people completed all stages of the study for a completion rate of 75\%. 
The 25 participants were made up of 4 undergraduate students (16\%), 19 graduate students (76\%), and 2 professionals (8\%). 
A majority of our participants (19/21) described themselves as ``first time users'' of Copilot, 5 of them indicated that they had ``tried it out a few times'' and 1 indicated that they ``used it all the time''.

\begin{table}
\begin{tabularx}{\linewidth}{|X|X|X|X|}
\hline
\textbf{CWE-ID} & \textbf{Description} & \textbf{Problem S} & \textbf{Problem T} \\ \hline
CWE-20                                & Improper Input Validation                 &\checkmark             &\checkmark                                         \\ \hline
CWE-22                                & Path Traversal                            &                                         &\checkmark                                         \\ \hline
CWE-78                                & OS Command Injection                      &                                         &\checkmark                                         \\ \hline
CWE-79                                & Cross-Site Scripting                      &\checkmark                                         &                                         \\ \hline
CWE-89                                & SQL Injection                             &\checkmark                                         &\checkmark                                         \\ \hline
CWE-125                               & Out of Bounds Read                        &\checkmark                                         &\checkmark                                        \\ \hline
CWE-285                               & Improper Authorization                    &\checkmark                                         &\checkmark                                         \\ \hline
CWE-287                               & Improper Authentication                   &\checkmark                                         &\checkmark                                         \\ \hline
CWE-401                               & Memory Leak                               &\checkmark                                         &\checkmark                                         \\ \hline
CWE-415                               & Double Free                               &\checkmark                                         &\checkmark                                         \\ \hline
CWE-416                               & Use After Free                            &\checkmark                                         &\checkmark                                         \\ \hline
CWE-476                               & Null Pointer Dereference                  &\checkmark                                         &\checkmark                                         \\ \hline
CWE-787                               & Out of Bounds Write                       &\checkmark                                         &\checkmark                                         \\ \hline
\end{tabularx}
\caption{The list of \glspl{cwe} that we specifically checked for in each problem. Problems were designed such that the specified \gls{cwe} could be introduced if participants were not careful enough about writing secure code.}
\label{table:tested-cwes}
\end{table}

\subsection{Material: Programming Problems}

\subsubsection{Problem Design}
We designed two problems for this study: problem S and problem T.
In problem S, the participants had to implement a sign-in function for an application given a user's identifier and password. 
In problem T, the participants had to implement a function that performs a series of transactions in a given transaction file and then renames the file.

We decided to create our own set of problems for this study as we had specific criteria that called for tailored problems. Specifically, we sought problems that:

\begin{enumerate}
    \item\label{crite1} had potential for vulnerable solutions,
    \item\label{crite2} had solutions that could manually be analyzed
    \item\label{crite3} resembled real world applications
    \item\label{crite4} could be solved by participants within an hour
\end{enumerate}

In order to address criteria \ref{crite1} and \ref{crite2}, we designed the problems so that certain vulnerabilities could be introduced if participants were not careful with their solutions. The vulnerabilities that could be introduced were based on Common Weakness Enumerations (CWEs) \cite{noauthor_mitre_nodate}. \glspl{cwe} are weaknesses in software and hardware systems and they are defined and maintained by the Mitre corporation. A CWE essentially represents a vulnerability and can be identified by its unique ID. For example, the classic buffer overflow vulnerability is represented by CWE-120. In designing each problem, we curated a set of \glspl{cwe} that we wanted to include. We selected \glspl{cwe} either because they were included on Mitre's top 25 most dangerous software weaknesses list \cite{top_25_mitre} or because they were pertinent to the C programming language which was to be used in the study. The set of possible \glspl{cwe} for each problem would subsequently be used for our analysis of participant solutions. Table \ref{table:tested-cwes} contains the \glspl{cwe} that we focused on for each problem. Note that this set of \glspl{cwe} is not exhaustive and there could have been other vulnerabilities possible in the problems we designed.

In order to address criteria \ref{crite3} and \ref{crite4}, we created a collection of well documented helper functions (with stub implementations) for each problem. These helper functions enabled us to expand the level of difficulty of our problems (approximating real world applications) while also constraining and guiding users towards finding solutions within a confined solution space in the allotted time. We conducted pre-study testing to verify that Copilot could generate solutions using the stub helper functions provided in the problem file. After the study had commenced, it became apparent that there was a difference in difficulty between the two problems. We discuss this variation and how we adapted to it in section \ref{sec:results-overview}.

Other user studies conducted around the security of \glspl{cgt} have designed different types of problems to different ends. The study by Sandoval et al.\cite{sandoval_lost_2023} designed a single large problem in the C programming language that participants had to solve within two weeks. However, Perry et al.\cite{perry_users_2022} designed 6 (relatively) smaller problems in different programming languages including JavaScript and C that participants had to solve within 2 hours (20 minutes each). These problems were less open-ended and had more straightforward solutions.
 
\subsection{Procedure}\label{subsection:problem-scoring}

\subsubsection{Problem Solving}
We employed a within-subject study design wherein all participants served in both the treatment group and the control group. All participants solved one programming problem with Copilot access and the other problem without Copilot access.
Each participant involved in the study was programmatically assigned to one of four groups on a round-robin basis. The groups determined the order in which the problems were solved and whether Copilot would be used to solve problem S or problem T.
There were four groups because there were two variables that determined how participants would solve the problems, and each variable had two possibilities. Any given participant could solve problem S first or problem T first. At the same time, the participant could either solve the first problem with Copilot and the second problem without Copilot, or vice versa.
Participants were given 60 minutes to solve each problem together with an instruction sheet that they could reference during problem solving.
Participants were informed (verbally and in the written instructions) that they were to write secure code.
There were no restrictions on the resources participants could consult to aid in solving the problem other than the restrictions on Copilot use and the use of other \glspl{cgt}.
Each participant's screen was recorded during problem solving for subsequent analysis after the study session.
All participants used Copilot in the Microsoft Visual Studio Code text editor. Participants were free to use and interact with Copilot in whatever manner they preferred.
Solutions were saved once participants were done solving a given problem. Then, participants were required to fill out surveys to provide additional information about their perspective on the problem they just completed.
Upon completion of the study, each participant was compensated CAD50.00.

\subsubsection{Functionality Analysis}
We tested solutions for functionality requirements with two of types tests: \textbf{basic tests} and \textbf{advanced tests}.
Participants had access to the basic test during the study and had the option of testing their solutions on it if they desired.
They did not have access to the advanced test.
To perform the basic test, participants had to uncomment and run code provided for them in the \lstinline{main} function of the problem file.
The basic tests tested participant solutions on simple inputs, similar to what was described in the instructions. 
The advanced testing involved checking participant solutions on edge-case and more complex inputs such as null inner structs (problem S) and multiple transactions (problem T).

\subsubsection{Security Analysis}
All solutions submitted by participants were checked for the presence of the various \glspl{cwe} possible for each problem (Table \ref{table:tested-cwes}). This checking was performed manually by one author and one other independent coder. We resorted to manual analysis of participant solutions because it has been proven to be sufficient when it comes to analyzing relatively short snippets of code~\cite{pearce_asleep_2022, perry_users_2022, siddiq_securityeval_2022}. Other research that has performed security analyses of code generated by \glspl{cgt} has generally relied on manual analysis or CodeQL \cite{github_inc_codeql_2019} to check for the presence of vulnerabilities. We used manual analysis because our preliminary testing of CodeQL showed that it was unable to detect any of the vulnerabilities in our test samples - it always generated false negative results. To ensure that CodeQL's poor performance was not due to any misconfiguration of our CodeQL setup, we performed additional tests to validate its setup. We used code snippets provided in the CodeQL GitHub repository \cite{github_inc_codeql_2019-1} which were known to contain certain \glspl{cwe}. For these examples, CodeQL was able to successfully identify the vulnerabilities. We also considered using fuzzing for our analyses but decided against it due to the proven track record of manual analysis and the use of stub helper functions in the programming problems, which would make adopting fuzzing a costly endeavour with no guarantee of better performance.

Solutions were analyzed independently by one of the authors and an independent coder - a Computer Science PhD student in our department with experience in C programming and vulnerability analysis. 
For each solution, both parties checked for the presence of each of the \glspl{cwe} in Table \ref{table:tested-cwes} and stored their results separately. 
The results were subsequently cross-referenced to find situations where the author and coder were in disagreement about the presence or absence of a CWE in a solution. A vulnerability was only considered present or absent in a solution if both the author and the coder were in agreement.
Where there were disagreements, the author and the coder discussed until a consensus was reached. This was required in less than 5\% of the cases. The kappa score, a metric used to measure inter-rater reliability, was calculated to be 0.962 for the vulnerability labeling in this study, indicating that the level of agreement between the coder and the author was close to perfect.
The manual analysis of solutions resulted in a security score for the two problems solved by each participant. The security score, outlined below, is a function of the number of vulnerabilities present in a participant's solution. For our purposes, a solution with a higher security score is more secure than a solution with a lower security score. The security score ranges from 0 (all vulnerabilities found) to 100 (no vulnerabilities found). While the security score was computed for all solutions, only those that compiled and passed the basic test were used for subsequent analyses. 



\begin{equation*}
\resizebox{\linewidth}{!}{$Percentage Vulnerable = \frac{\text{Number of Vulnerabilities found}}{\text{Total number of Vulnerabilities Possible}} * 100$}
\end{equation*}

\begin{equation*}
\resizebox{\linewidth}{!}{$\textbf{Security Score} = 100 - Percentage Vulnerable$}
\end{equation*}

\begin{table*}
\centering
\begin{tabular}{|c|ccc|ccc|ccc|}
\hline
               & \multicolumn{3}{c|}{\textbf{Educational Level}}                                   & \multicolumn{3}{c|}{\textbf{Dev. Experience (Yrs)}} & \multicolumn{3}{c|}{\textbf{Copilot Experience}}                                                     \\ \hline
               & \multicolumn{1}{c|}{Undergraduate} & \multicolumn{1}{c|}{Graduate} & Professional & \multicolumn{1}{c|}{1-5}  & \multicolumn{1}{c|}{6-10}  & greater than 10  & \multicolumn{1}{c|}{``first time user''} & \multicolumn{1}{c|}{``tried it out''} & ``frequent user'' \\ \hline
\textbf{Count} & \multicolumn{1}{c|}{4}             & \multicolumn{1}{c|}{19}       & 2            & \multicolumn{1}{c|}{14}   & \multicolumn{1}{c|}{8}     & 3   & \multicolumn{1}{c|}{19}                  & \multicolumn{1}{c|}{5}                & 1                 \\ \hline
\end{tabular}
\caption{Summary of participants' background}
\label{table:participant-background}
\end{table*}

\lstset{style=mystyle}

\subsection{Ethics}

This user study obtained ethics clearance from the Human Research Ethics Board at the University of Waterloo.
Participant consent was obtained during the recruitment process and consenting participants were screened to ensure they met the desired criteria.
Participants were informed that their screens would be recorded during the session.
Data collected during sessions, including screen recordings, problem solutions, and survey information, were linked to anonymous IDs created for each participant. We maintained a key in a secure vault linking participant information (name and email address) to IDs that will be deleted once all analysis is complete and no further contact with participants is required.

\section{Results and Discussion}\label{sec:results}

\subsection{Overview}\label{sec:results-overview}

\begin{table*}
\begin{tabularx}{\textwidth}{|p{0.05\textwidth}|X|X|X|X|X|X|}
\hline
\textbf{ID}  & \textbf{PS Score}                      & \textbf{PT Score}                      & \textbf{PS Time (mins)} & \textbf{PT Time (mins)} & \textbf{PS Func.} & \textbf{PT Func.} \\ \hline
001 & \cellcolor{cellcolor}54.5 & 41.7 & 23 & 28 & 3 & 2  \\ \hline
002 & 81.8 & \cellcolor{cellcolor}58.3 & 25 & 23 & 2 & 2  \\ \hline
003 & 81.8 & \cellcolor{cellcolor}91.7 & 42 & 51 & 3 & 3  \\ \hline
004 & \cellcolor{cellcolor}81.8 & 50.0 & 14 & 38 & 3 & \cellcolor{b_color}1  \\ \hline
005 & \cellcolor{cellcolor}63.6 & 41.7 & 10 & 46 & 3 & 2  \\ \hline
006 & 63.6 & \cellcolor{cellcolor}41.7 & 27 & 41 & 3 & 2  \\ \hline
007 & 81.8 & \cellcolor{cellcolor}50.0 & 20 & 25 & 3 & 3  \\ \hline
008 & \cellcolor{cellcolor}90.9 & 41.7 & 14 & 35 & 3 & 3  \\ \hline
009 & \cellcolor{cellcolor}72.7 & 58.3 & 13 & 59 & \cellcolor{b_color}1 & \cellcolor{b_color}1  \\ \hline
010 & 81.8 & \cellcolor{cellcolor}33.3 & 50 & 53 & 3 & 3  \\ \hline
011 & 100.0 & \cellcolor{cellcolor}50.0 & 60 & 40 & \cellcolor{b_color}0 & \cellcolor{b_color}1  \\ \hline
012 & \cellcolor{cellcolor}81.8 & 33.3 & 38 & 60 & 3 & \cellcolor{b_color}1  \\ \hline
013 & \cellcolor{cellcolor}81.8 & 41.7 & 11 & 55 & 2 & 2  \\ \hline
014 & 81.8 & \cellcolor{cellcolor}75.0 & 25 & 21 & 3 & 3  \\ \hline
015 & 63.6 & \cellcolor{cellcolor}66.7 & 60 & 19 & \cellcolor{b_color}1 & 2  \\ \hline
016 & \cellcolor{cellcolor}81.8 & 41.7 & 29 & 47 & 3 & 3  \\ \hline
017 & \cellcolor{cellcolor}36.4 & 25.0 & 25 & 60 & \cellcolor{b_color}1 & \cellcolor{b_color}1  \\ \hline
018 & 90.9 & \cellcolor{cellcolor}50.0 & 26 & 51 & 3 & 3  \\ \hline
019 & 72.7 & \cellcolor{cellcolor}66.7 & 21 & 22 & 3 & 3  \\ \hline
020 & \cellcolor{cellcolor}81.8 & 50.0 & 27 & 39 & 2 & 2  \\ \hline
021 & \cellcolor{cellcolor}81.8 & 91.7 & 15 & 42 & 3 & 3  \\ \hline
022 & 90.9 & \cellcolor{cellcolor}33.3 & 21 & 32 & 3 & 3  \\ \hline
023 & 81.8 & \cellcolor{cellcolor}50.0 & 41 & 60 & 3 & \cellcolor{b_color}1  \\ \hline
024 & \cellcolor{cellcolor}81.8 & 66.7 & 11 & 60 & \cellcolor{b_color}0 & \cellcolor{b_color}0  \\ \hline
025 & \cellcolor{cellcolor}72.7 & 50.0 & 35 & 60 & 2 & 2  \\ \hline
\end{tabularx} 
\caption{Participants in the study and their performance (security scores) on problem S (PS) and problem T (PT). \colorbox{cellcolor}{Highlighted} cells in the ``PS Score'' and ``PT Score'' columns indicate that the score was obtained with Copilot. The time columns show the times taken to solve each problem. The last two functionality columns indicate the level of functionality of participant solutions which are described as follows: 0 = did not compile, 1 = only compiled, 2 = compiled and passed only the basic test, 3 = compiled and passed both the basic and the advanced test. \colorbox{b_color}{Highlighted} cells in the functionality columns indicate solutions that did not sufficiently implement the functionality requested in the problems and were therefore excluded from our analysis.}
\label{table:all-results}
\end{table*}

Table \ref{table:participant-background} summarizes the demographic data that we collected about participants, namely their level of education, experience writing code and experience with Copilot.
Table \ref{table:all-results} summarizes the data about participant performance in our study. 17 participants submitted valid solutions for both problems. As mentioned earlier, valid solutions were those that compiled and at least passed the basic test. In table \ref{table:all-results}, valid solutions for a problem correspond to rows where PS Func. or PT Func. are greater than or equal to 2 (highlighted in yellow). Of the 17 that submitted valid solutions for both problems, 8 were better with Copilot (i.e. wrote more secure code) and 9 were better without Copilot. The average security score with Copilot (\textbf{65.2}, std=18.5) was lower than the average security score without Copilot (\textbf{66.3}, std=19.8).

Looking at the problems separately, 20 participants submitted valid solutions for problem S and 18 participants submitted valid solutions for problem T. On average, participants took 27.4 minutes to submit a solution for problem S (std=14.1) and 42.7 minutes to submit a solution for problem T (std=13.8).

Before the study officially began, we tested our problems on two volunteers who were representative of the kind of people we expected to be in the actual study. These volunteers provided feedback which we used to edit our problems before proceeding with study. The feedback they provided addressed two points: the high level of difficulty of one of the problems (problem T) and the clarity of some of the instructions. While we attempted to address both concerns, our results (time taken and average security scores) indicate that the comparatively higher difficulty of one of the problems over the other may have persisted. As a result, we adapted our analysis and discussion to account for this variation in difficulty. We specifically looked at the results from solving each problem and performed statistical tests independently. All calculations and data aggregations as well as visualizations were duplicated for both problems.

\subsection{Copilot Suggestion Analysis}


To understand the extent to which Copilot contributed to participants' solutions, we performed some analysis of the screen recordings generated during study sessions with participants. We were unable to track the provenance (i.e., from the participant or from Copilot) of each vulnerability found through this process because participants accepted and edited code, and interacted with Copilot in several different ways. 
In most cases, the presence or absence of a vulnerability could only be determined when the participant had finished editing the file.
Thus, while we could know that Copilot had suggested a piece of code, we could not at all times check for vulnerabilities because these vulnerabilities existed within the larger context of the file.
For this analysis, we focused on tracking the number of suggestions Copilot made, the number of suggestions that were accepted, and the number of suggestions that were edited after being accepted. 

Our analysis of Copilot suggestions yielded two important takeaways. First, we noticed that the acceptance rate did not change
significantly between the two problems, despite their different difficulty levels. Participants used it at the same rate for the easier
problem (problem S) as they did for the harder problem (problem T). The second takeaway was that Copilot played at least a minor
role in all solutions submitted by participants for problems where Copilot was permitted. This is evident from the fact that all participants accepted at least 6 Copilot suggestions, each with at least 1 line of code. Table \ref{table:copilot-suggestions} provides a quantitative summary of the Copilot suggestion analysis.

\begin{table*}[]
\begin{tabular}{|c|ccccc|clcl|}
\hline
                   & \multicolumn{5}{c|}{\textbf{Number of Copilot Suggestions}}                                                                                                                                                                                                                           & \multicolumn{4}{c|}{\textbf{Participant Reaction}}                                                                                                         \\ \hline
                   & \multicolumn{1}{c|}{\textbf{Mean}} & \multicolumn{1}{c|}{\textbf{Median}} & \multicolumn{1}{c|}{\textbf{Std. Dev.}} & \multicolumn{1}{c|}{\textbf{\begin{tabular}[c]{@{}c@{}}Range\\ (Suggested)\end{tabular}}} & \textbf{\begin{tabular}[c]{@{}c@{}}Range\\ (Accepted)\end{tabular}} & \multicolumn{1}{c|}{\textbf{Avg. Acc. Rate}} & \multicolumn{1}{l|}{\textbf{Std. Dev.}} & \multicolumn{1}{c|}{\textbf{Avg. Edit Rate}} & \textbf{Std. Dev.} \\ \hline
\textbf{Problem S} & \multicolumn{1}{c|}{10.90}         & \multicolumn{1}{c|}{11}              & \multicolumn{1}{c|}{6.23}               & \multicolumn{1}{c|}{8 - 14}                                                               & 6 - 13                                                              & \multicolumn{1}{c|}{84.99\%}                 & \multicolumn{1}{l|}{11.17\%}            & \multicolumn{1}{c|}{22.42\%}                 & 26.12\%            \\ \hline
\textbf{Problem T} & \multicolumn{1}{c|}{21.10}         & \multicolumn{1}{c|}{21}              & \multicolumn{1}{c|}{1.79}               & \multicolumn{1}{c|}{14 - 29}                                                              & 8 - 24                                                              & \multicolumn{1}{c|}{83.53\%}                 & \multicolumn{1}{l|}{12.61\%}            & \multicolumn{1}{c|}{17.16\%}                 & 12.81\%            \\ \hline
\textbf{Both}      & \multicolumn{1}{c|}{16.00}         & \multicolumn{1}{c|}{14}              & \multicolumn{1}{c|}{4.58}               & \multicolumn{1}{c|}{8 - 29}                                                               & 6 - 24                                                              & \multicolumn{1}{c|}{84.26\%}                 & \multicolumn{1}{l|}{11.62\%}            & \multicolumn{1}{c|}{19.79\%}                 & 20.21\%            \\ \hline
\end{tabular}
\caption{Describing the nature of Copilot suggestions and how users interacted with it. The table shows the mean, median, and the range of the number of suggestions made by Copilot as well as the range of the number of accepted suggestions and the acceptance and edit rates.}
\label{table:copilot-suggestions}
\end{table*}

\subsection{\textbf{RQ1}: Does Copilot use correlate with participants writing more secure code?}\label{sec:rq1}

\subsubsection{Approach}
To investigate the possible effects of Copilot on the security of participant solutions, we looked at participant security scores with and without Copilot. 
We first computed summary statistics (mean, median, standard deviation) of security scores for both problems. This gave as an overview of the overall performance (per problem) with and without the assistance of Copilot. 
We subsequently performed statistical tests to see whether there was a significant difference between the two groups.
For each problem, we used the Kruskal-Wallis to test for statistically significant differences between the group that used Copilot and the group that did not. This test was performed independently for each problem to account for any differences in their level of difficulty. We chose the Kruskal-Wallis test because it allowed us to compare the scores from the two independent groups even when the data did not follow a normal distribution, an assumption made by other (parametric) tests like the T-test. 

\begin{figure}
    \centering
    \begin{subfigure}{\linewidth}
        \includegraphics[width=\linewidth]{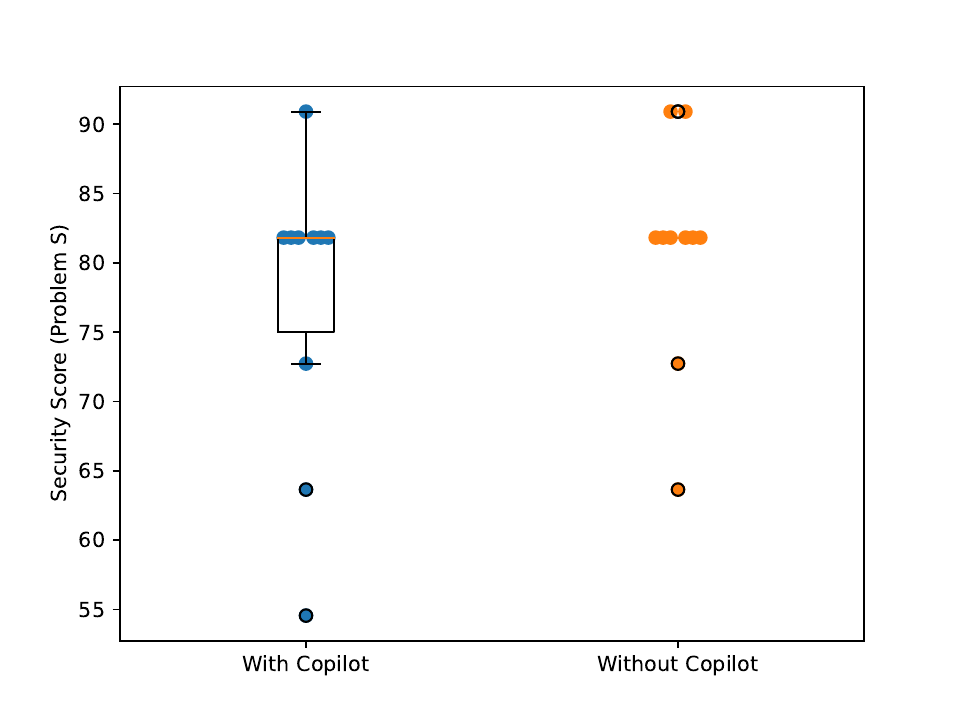}
        \caption{Box plots describing security scores with and without the use of Copilot for Problem S.}
        \label{fig:ps-box}
    \end{subfigure}
    \begin{subfigure}{\linewidth}
        \centering
        \begin{tabular}{|c|c|c|c|}
        \hline
        & \textbf{Mean} & \textbf{Median} & \textbf{Std. Dev.} \\
        \hline
        \textbf{With Copilot} & 77.27 & 81.82 & 10.71 \\
        \hline
        \textbf{Without Copilot} & 80.91 & 81.82 & 7.96 \\
        \hline
        \end{tabular}
        \caption{Descriptive statistics of participant's performance for Problem S.}
        \label{table:ps-stats}
    \end{subfigure}
    \caption{Box plot and table summarizing participant's performance for Problem S with and without the use of Copilot.}
    \label{fig:ps-overall}
\end{figure}

\begin{figure}
    \centering
    \begin{subfigure}{\linewidth}
        \includegraphics[width=\linewidth]{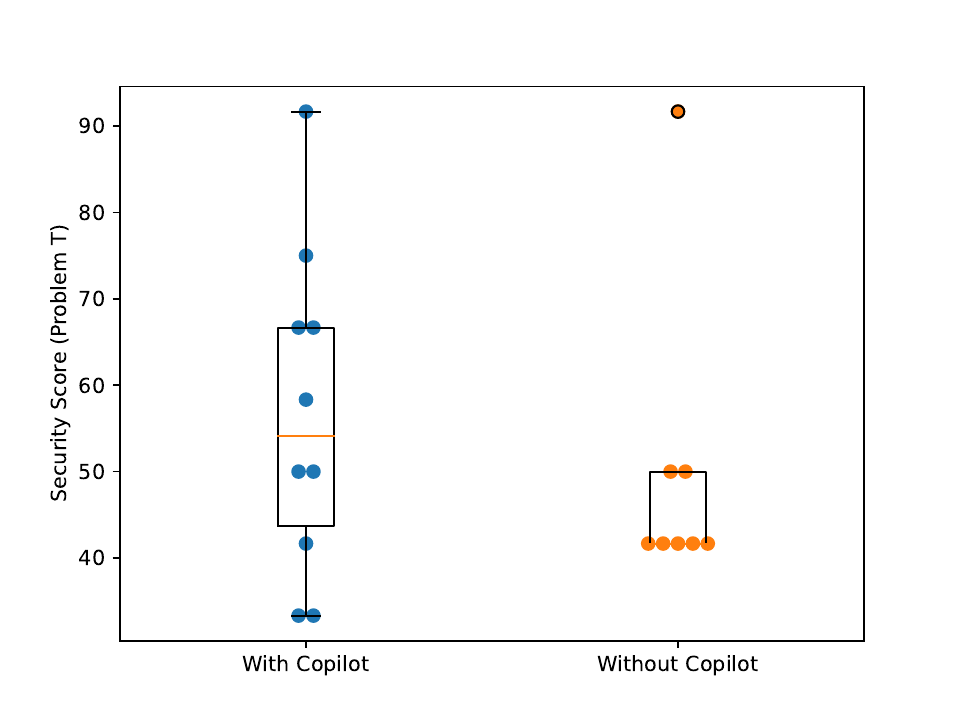}
        \caption{Box plots describing security scores with and without the use of Copilot for Problem T.}
        \label{fig:pt-box}
    \end{subfigure}
    \begin{subfigure}{\linewidth}
        \centering
        \begin{tabular}{|c|c|c|c|}
        \hline
        & \textbf{Mean} & \textbf{Median} & \textbf{Std. Dev.} \\
        \hline
        \textbf{With Copilot} & 56.67 & 54.17 & 18.76 \\
        \hline
        \textbf{Without Copilot} & 50.00 & 41.67 & 17.25 \\
        \hline
        \end{tabular}
        \caption{Descriptive statistics of participant's performance for Problem T.}
        \label{table:pt-stats}
    \end{subfigure}
    \caption{Box plot and table summarizing participant's performance for Problem T with and without the use of Copilot.}
    \label{fig:pt-overall}
\end{figure}

\subsubsection{Results}
We used the security score (computed following the steps in subsection \ref{subsection:problem-scoring}) as the basis for evaluating the security of solutions. Figure \ref{fig:ps-overall} summarizes the impacts of the use of Copilot on security scores for problem S and Figure \ref{fig:pt-overall} does the same for problem T. For problem S, we obtained the same median security score of 81.82 both for participants who solved it with Copilot access and for participants who solved it without Copilot access. For problem T, the median security score for participants who solved it with Copilot access was 54.17 compared to 41.67 for participants who solved it without Copilot. 


Using the Kruskal-Wallis statistical test, we found no statistically significant differences in security scores for both problem S (statistic = 0.59, p = 0.44) and problem T (statistic = 0.83, p = 0.36). The results of the tests indicate that we cannot reject the possibility that Copilot has no effect on the security of code written by participants.

However, from our sample, looking specifically at problem T, we observe a marked difference between the median security score with Copilot and median security score without Copilot - the score with Copilot is higher by about 13 points. This difference in scores also applies to the mean; the score with Copilot is higher than the score without Copilot by about 6 points. On the other hand, we see no such differences in scores for problem S - the medians are exactly the same. Considering that problem T appeared to be more difficult for participants to solve (it took longer to solve on average and had lower security scores overall), it seems that Copilot benefited participants when they encountered the more complex problem and had little effect when the problem was more straightforward. 

A possible explanation for this difference in performance is that when presented with the harder problem, participants' priorities shifted from finding \textit{a secure solution} to finding \textit{any solution}. To achieve this, participants may have been less concerned about the security of the code they were writing. Those who had access to Copilot for this problem may also have been less concerned with the level of security of Copilot suggestions as indicated by the lower edit rate for problem T in table \ref{table:copilot-suggestions}. However, even if the participant's priorities had changed, Copilot's \textit{priorities} remained the same. Under these circumstances, participants who had access to Copilot for problem T benefited from its ability to not sacrifice security for expediency or functionality. The flip side of this discussion, which we cannot verify from the perspective of this study, is that since Copilot's priorities remain unchanged, users who prioritize security at least as much as functionality may be negatively impacted by using it.


\noindent\fbox{%
    \parbox{\linewidth}{%
        \textbf{Summary:} In RQ1, we investigated whether using Copilot correlates with participants writing more secure code. While not statistically significant, we observed that participants wrote more secure code when they had access to Copilot for the more difficult problem.
    }%
}

\subsection{\textbf{RQ2}: Are there vulnerability types that Copilot is more susceptible to or more resilient against?}\label{sec:user_vuln_analysis}\label{sec:rq2}

\subsubsection{Approach}
We investigated the possibility of Copilot having a disproportionate impact on certain vulnerability types by looking at the frequency of vulnerabilities and how that frequency changed with and without the use of Copilot. We further ran Fisher's exact statistical test on the collected counts to determine whether Copilot's impact on the presence/absence of a vulnerability was statistically significant. Tests on Copilot's impact on the different vulnerabilities were performed separately for each problem, but we also performed a joint analysis for vulnerabilities that were common to both problems.

\begin{table*}
\small
\begin{tabular}{|c|cc|cc|ccc|}
\hline
                 & \multicolumn{2}{c|}{\textbf{Problem S}}                                 & \multicolumn{2}{c|}{\textbf{Problem T}}                                 & \multicolumn{3}{c|}{\textbf{Total}}                                                                                    \\ \hline
                 & \multicolumn{1}{c|}{\textit{\textbf{With}}} & \textit{\textbf{Without}} & \multicolumn{1}{c|}{\textit{\textbf{With}}} & \textit{\textbf{Without}} & \multicolumn{1}{c|}{\textit{\textbf{With}}} & \multicolumn{1}{c|}{\textit{\textbf{Without}}} & \textit{\textbf{Total}} \\ \hline
\textbf{CWE-20}  & \multicolumn{1}{c|}{7}                      & 5                         & \multicolumn{1}{c|}{9}                      & 7                         & \multicolumn{1}{c|}{16}                     & \multicolumn{1}{c|}{12}                         & 28                      \\ \hline
\textbf{CWE-22}  & \multicolumn{1}{c|}{-}                      & -                         & \multicolumn{1}{c|}{4}                      & 7                         & \multicolumn{1}{c|}{4}                      & \multicolumn{1}{c|}{7}                         & 11                       \\ \hline
\textbf{CWE-78}  & \multicolumn{1}{c|}{-}                      & -                         & \multicolumn{1}{c|}{5}                      & 7                         & \multicolumn{1}{c|}{5}                      & \multicolumn{1}{c|}{7}                         & 12                      \\ \hline
\textbf{CWE-79}  & \multicolumn{1}{c|}{3}                      & 2                         & \multicolumn{1}{c|}{-}                      & -                         & \multicolumn{1}{c|}{3}                      & \multicolumn{1}{c|}{2}                         & 5                       \\ \hline
\textbf{CWE-89}  & \multicolumn{1}{c|}{2}                      & 1                         & \multicolumn{1}{c|}{9}                      & 7                         & \multicolumn{1}{c|}{11}                      & \multicolumn{1}{c|}{8}                         & 19                      \\ \hline
\textbf{CWE-125} & \multicolumn{1}{c|}{0}                      & 0                         & \multicolumn{1}{c|}{1}                      & 0                         & \multicolumn{1}{c|}{1}                      & \multicolumn{1}{c|}{0}                         & 1                       \\ \hline
\textbf{CWE-285} & \multicolumn{1}{c|}{1}                      & 1                         & \multicolumn{1}{c|}{1}                      & 2                         & \multicolumn{1}{c|}{2}                      & \multicolumn{1}{c|}{3}                         & 5                       \\ \hline
\textbf{CWE-287} & \multicolumn{1}{c|}{0}                      & 1                         & \multicolumn{1}{c|}{4}                      & 2                         & \multicolumn{1}{c|}{4}                      & \multicolumn{1}{c|}{3}                         & 7                       \\ \hline
\textbf{CWE-401} & \multicolumn{1}{c|}{10}                      & 9                         & \multicolumn{1}{c|}{10}                      & 8                         & \multicolumn{1}{c|}{20}                     & \multicolumn{1}{c|}{17}                        & 37                      \\ \hline
\textbf{CWE-415} & \multicolumn{1}{c|}{0}                      & 0                         & \multicolumn{1}{c|}{0}                      & 0                         & \multicolumn{1}{c|}{0}                      & \multicolumn{1}{c|}{0}                         & 0                       \\ \hline
\textbf{CWE-416} & \multicolumn{1}{c|}{0}                      & 0                         & \multicolumn{1}{c|}{0}                      & 0                         & \multicolumn{1}{c|}{0}                      & \multicolumn{1}{c|}{0}                         & 0                       \\ \hline
\textbf{CWE-476} & \multicolumn{1}{c|}{2}                      & 2                         & \multicolumn{1}{c|}{7}                      & 7                         & \multicolumn{1}{c|}{9}                      & \multicolumn{1}{c|}{9}                         & 18                      \\ \hline
\textbf{CWE-787} & \multicolumn{1}{c|}{0}                      & 0                         & \multicolumn{1}{c|}{2}                      & 1                         & \multicolumn{1}{c|}{2}                      & \multicolumn{1}{c|}{1}                         & 3                       \\ \hline
\end{tabular}
\caption{Counts of the number of times each \gls{cwe} was found for both problems. The ``With'' columns indicate the number of times a \gls{cwe} was found when Copilot was involved in solving the problem. ``Without'' indicates that Copilot was not involved. Dashes indicate that the particular \gls{cwe} was not tested for in that problem.}
\label{table:vulnerability-counts}
\end{table*}

\subsubsection{Results}
Table \ref{table:vulnerability-counts} presents the data about the different vulnerabilities found for each problem with and without the use of Copilot. For problem S, a total of 46 vulnerabilities were found. 25 were found with Copilot (i.e. were found when the participant was allowed to use Copilot) and 21 were found without Copilot. Overall, 54\% of vulnerabilities were found with Copilot for problem S. For problem T, a total of 100 vulnerabilities were found, with 52 (56\%) being found with Copilot and the remaining 48 without. An inspection of these summary statistics and the frequencies of each individual \gls{cwe} with and without Copilot did not reveal any clear or significant impact of Copilot on the presence of any particular vulnerability. To be sure, we also ran Fisher's exact test on a 2x11 contingency table for problem S and a 2x12 contingency table for problem T using the frequencies in Table \ref{table:vulnerability-counts} as the counts. 
The results of the tests for both problems indicated that there was no statistically significant difference between frequencies with Copilot and frequencies without Copilot (p=0.99 for problem S, p=0.94 for problem T). The results further indicate that as far as our sample is concerned, we cannot reject the possibility that Copilot has any statistically significant effect on the presence of the \glspl{cwe} tested in this study. 

\noindent\fbox{%
    \parbox{\linewidth}{%
        \textbf{Summary:} In RQ2, we investigated Copilot's influence on the presence of certain \glspl{cwe}. We found no significant impact of Copilot access on the introduction of the \glspl{cwe} we considered.
    }%
}

\subsection{Survey Results}

\begin{figure}
    \centering
        \includegraphics[width=\linewidth]{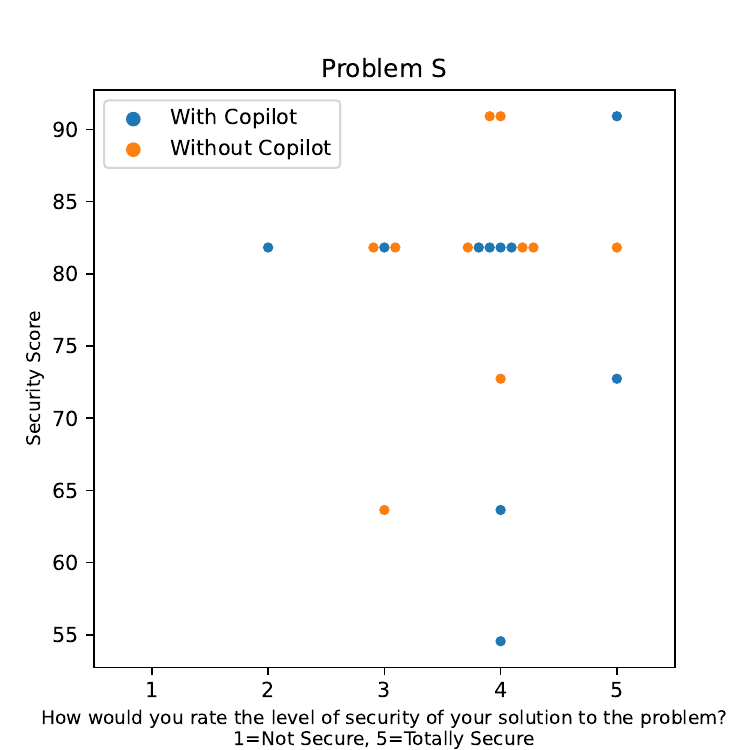}
        \caption{Plot showing how participants' opinions compared to their security scores with and without Copilot for problem S.}
    \label{fig:ps-opinions}
\end{figure}

\begin{figure}
    \centering
        \includegraphics[width=\linewidth]{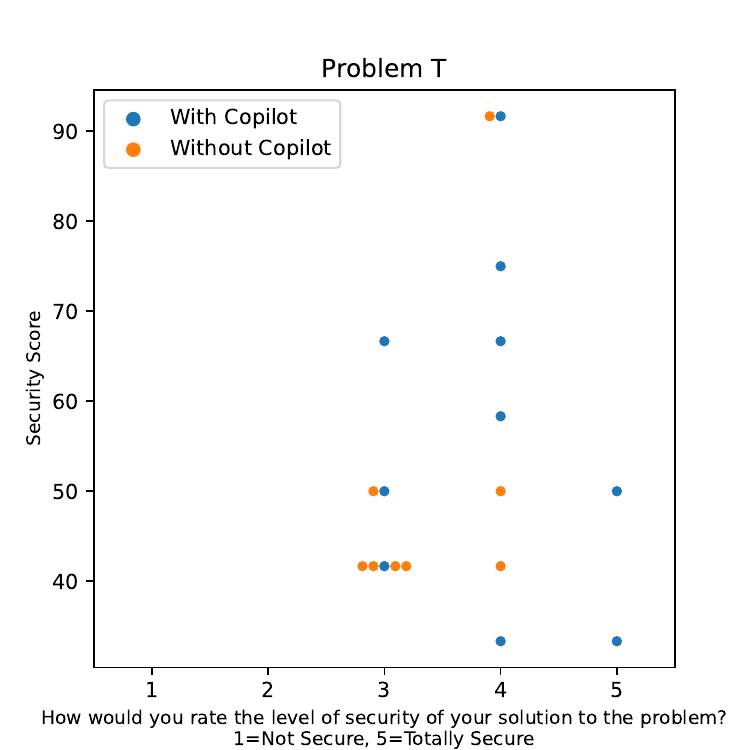}
        \caption{Plot showing how participants' opinions compared to their security scores with and without Copilot for problem T.}
    \label{fig:pt-opinions}
\end{figure}

After solving each problem, participants were asked to fill out surveys.
The amount of time that participants used to solve each problem was also recorded. For both problems, we found that the median time used in implementing a solution was less with Copilot than without. This is not surprising considering the high suggestion acceptance rates for both problems (Table \ref{table:copilot-suggestions}). When we asked participants to rate how helpful Copilot had been on a scale of 1 (not helpful) to 5 (very helpful), 64\% of them (16 out of 25) indicated that Copilot had been very helpful, giving it the maximum rating of 5. Of the remaining participants, 6, 2, and 1 of them rated Copilot's helpfulness at a 4, 3, and 2 respectively.

We also asked participants to provide ratings on how secure they felt their solutions were and how confident they were that their solutions were correct. We observed that opinions on correctness and security were generally high with a majority of participants giving ratings of 4 or 5 (out of 5) on both accounts. However, this was more true for correctness where high scores were given 82\% of the time than for security where high scores were given 58\% of the time.  

Figures \ref{fig:ps-opinions} and \ref{fig:pt-opinions} show how participant opinions on the level of security of their submissions compared to their actual security scores for problem S and problem T respectively. For problem S, we observed no significant trends between participant opinions and their corresponding security scores. For problem T, we observed that participants without Copilot were generally not as confident in the security of their solution as those with Copilot access and this low confidence loosely tracked with the lower security scores in this group as previously discussed.

\subsection{Takeaways \& Future work}
In this study, we investigated the the impact of using Copilot on code security (RQ1) and the amount of influence that Copilot has on the presence of certain CWEs (RQ2). While our findings are not statistically significant due to a limited sample size, for the former, we observed that participants wrote more code secure for the more difficult problem when they were granted access to Copilot. In the latter, we observed no significant impact of Copilot access on the presence of any of the CWEs considered in this study.

In order to verify whether Copilot does in fact make a significant difference on harder problems as suggested by the result from RQ1, a more targeted user study may be required. Such a study would require a set of multiple problems, each with varying levels of difficulty. Ideally, we would want to establish a proxy for problem difficulty that can be compared to a proxy for security (like the security score in this user study) during analysis. The proxy for problem difficulty could be obtained from a number of sources including the time taken to solve problems or some aggregate of rankings of problem difficulty by users after they have solved the problems. Participants in the study would then be split into control and treatment groups, wherein the former solve all problems without Copilot and the latter solve all the problems with Copilot. We would then be able to measure how differences in security performance between the control and treatment group are affected by problem difficulty.

The idea that the security impacts of Copilot could be more significant at higher levels of difficulty has implications for future research on and testing of Copilot and similar \glspl{cgt}. Mainly, it implies that testing \glspl{cgt} on trivial problems could yield misleading results. Researchers and developers of \glspl{cgt} may want to take steps to ensure that the problems upon which their tools are tested and evaluated are above a certain level of complexity, especially when the tools are being tested in conjunction with human users. For regular users of Copilot, a takeaway from the observations from our sample is the suggestion that Copilot can be especially helpful in writing more secure code when tackling more complex problems.

\section{Threats to Validity}\label{sec:threats}

\subsection{Construct Validity}

A possible threat to the construct validity of this study is the manual analysis used to evaluate participant solutions. In order to check for the presence of vulnerabilities, we manually analyzed participant solutions. It is possible that this analysis process may have missed (false negative) or misidentified (false positive) certain vulnerabilities. As a mitigation, we used two evaluators to decrease the chances of false results.

Our participant screening process did not take participants' security experience into account. This was because we wanted to investigate the impact of Copilot use on code security regardless developers' of security experience. However, we believe a study that explores the relationship between the security experience of users of \glspl{cgt} and the security of code generated with \glspl{cgt} would be interesting and we leave it as an avenue for future work.

\subsection{External Validity}

Threats to the external validity of this study are the sample size and sample composition. While we observe some effects of Copilot on the security of solutions for problem T, the tests we perform indicate that our findings are not statistically significant. This indicates that we cannot assume that the observations in our sample generalize to the larger population. Further, the majority of our sample (approximately 90\%) comprised students, both at the graduate and undergraduate level. As a result, our observations may also not be generalizable to professional, full-time software developers. We relaxed our selection criteria and designed accessible problems in order to be able to reach a wider audience while retaining the integrity of the study. We also provided compensation for participants who completed the study. However, there were also time constraints that determined when we could no longer accept participants. For future studies, the goal would be to have the study open for a longer time and take additional steps to reach a wider audience outside of the university environment. Still, we believe the findings of this study are still useful. The study highlights potential security benefits of CGTs like Copilot, specifically their being more beneficial (security-wise) when solving difficult problems as well as their not being disproportionately susceptible to the CWEs considered. The study also sets the stage and direction for future studies on the security of CGTs and the ways in which they can be improved.
\section{Related Work}\label{sec:related}

\begin{table*}
\small
\begin{tabularx}{\textwidth}{|X|X|X|X|}
\hline
    & \textbf{This Study} & \textbf{Sandoval et al.~\cite{sandoval_lost_2023}} & \textbf{Perry et al.~\cite{perry_users_2022}}\\ 
\hline
\textbf{Tool Evaluated} & Copilot & Codex (code-cushman-001) & Codex (code-davinici-002)\\ 
\hline
\textbf{Sample Size} & 25 & 58 & 47\\ 
\hline
\textbf{Sample make-up} & CS Students and Professionals & CS Students & CS Students and Professionals\\
\hline
\textbf{Number of Problems} & 2 & 1 (subdivided into 12 functions) & 6\\
\hline
\textbf{Time Given} & 1 hour per problem & 2 weeks & 2 hours total\\
\hline
\textbf{Programming Languages}  & C & C & Python, JavaScript, C\\
\hline
\textbf{Problem Design} 
&
Participants were tasked with solving two problems: one that implemented user sign on a website and the other that implemented transaction fulfillment. In addition to other criteria, the problems were designed to mimic real world functionality, to be solvable within an hour, and to have the potential for insecure solutions. 
& 
Participants were asked to implement a shopping list based on a singly linked list data structure. The problem was designed to have the potential for several memory related bugs. 
&
Participants were asked to solve 6 relatively short problems in different languages. The problems were more direct in terms of security risks. Potential security risks were not obscured by higher level functionality requirements such as a shopping list or user sign in. For example participants were directly asked to implement cryptographic encryption,  message signing, and displaying a string input in a browser.\\
\hline

\textbf{Study Approach}
&
\textbf{All participants solved both problems} - one problem was solved with Copilot and the other problem without Copilot. This way, \textbf{each participant served in the treatment group for one problem and the control group for the other problem}.
&
Each participant was assigned to either the treatment or the control group.                                                                                                        & 
Each participant was assigned to either the treatment or the control group.\\
\hline
\textbf{Mode of CGT Use}
&
Participants used the Copilot extension in the Visual Studio Code text editor.
&
Participants used a custom VS Code extension connected to a codex model.
&
In addition to a custom UI for writing solutions, participants were provided a separate interface where they could query the codex model and then copy and paste results into their solution.\\
\hline

\textbf{Main Security Findings}
&
Participants generally submitted more secure solutions when they had access to AI assistance for the harder problem. For the easier problem, no difference was observed. We also observed no significant difference in performance across the different vulnerability types.
&
In their context, the LLM did not increase the incidence rate of severe vulnerabilities.
&
Participants with access to AI assistance produced more security vulnerabilities and were more likely to believe that they wrote secure code.\\
\hline

\textbf{CWEs Tested}
&
CWE-20, CWE-22, CWE-78, CWE-79, CWE-89, CWE-125, CWE-285, CWE-287, CWE-401, CWE-415, CWE-416, CWE-476, CWE-787
&
CWE-119, CWE-400, CWE-416, CWE-476, CWE-787, CWE-190, CWE-252, CWE-758, CWE-835
&
Did not make use of the CWE framework.\\
\hline

\end{tabularx}
\caption{Table summarizing the differences between three user studies, by different authors, on the effects of \glspl{cgt} on code security.}
\label{table:user-study-summary}
\end{table*}

There are a number of papers that have conducted evaluations of \glspl{cgt}~\cite{brown_language_2020, chen_evaluating_2021, li_competition-level_2022, xu_systematic_2022, ciniselli_what_2022, yan_whygen_2022, sobania_choose_2022, vaithilingam_expectation_2022, barke_grounded_2022, ziegler_productivity_2022}. As mentioned earlier, most evaluations tend to focus less on security. There are however two existing user studies of \glspl{cgt} with a focus on security that we would like to discuss briefly.

Sandoval et al. \cite{sandoval_lost_2023} conducted a user study that sought to investigate the cybersecurity impact of \glspl{llm} on code written by student programmers. They specifically evaluated the Codex language model on a sample size of 58 students. They found a small impact of \glspl{llm} on code security and a beneficial impact on functional correctness, indicating their use did not introduce new security risks but helped participants generate more correct solutions. 

On the other hand, Perry et al. \cite{perry_users_2022} also performed a large-scale study that also aimed to determine whether users wrote more insecure code with AI assistants. They also performed their evaluation using the Codex model and a sample size of 47 participants. They found that participants who had access to the Codex assistants wrote significantly less secure code than those without access, and were also more likely to believe they wrote more secure code.

An insight from our user-centered evaluation that is not present in the other studies is the idea that Copilot could be more beneficial (security-wise) for more difficult problems. Beyond that, we observe that our findings about \gls{cgt} security performance align slightly with those of Sandoval et al.~\cite{sandoval_lost_2023} in the sense that they both report either neutral or positive impacts of \glspl{cgt} on security. These studies have other things in common that could explain this similarity, specifically the focus on a single language (C) and the use of more in depth problems. On the other hand, we note that the findings by Perry et al.~\cite{perry_users_2022} tell a different story - indicating that \glspl{cgt} negatively impact the security performance of users. The simplest reason for this contradictory finding is the several differences in approach/methodology outlined in table \ref{table:user-study-summary}, chief among them being the fact that each study evaluates a different tool. The difference in results across the studies suggests that we may not want to generalize the performance of one \gls{cgt} to all other \glspl{cgt}. 
\section{Conclusion}\label{sec:conclusion}

In this user-centered evaluation of Copilot, we aimed to determine whether using Copilot correlated with participants writing more secure code (RQ1) and whether there were vulnerability types that Copilot was more susceptible to or more resilient against (RQ2). 
For RQ1, while there were no major differences in security performance between the two groups (with and without Copilot access) for problem S, we observed that the group with Copilot access for problem T (the relatively harder problem) tended to have higher security scores compared to the group without Copilot access for the same problem. We believe this may be due to the fact that when presented with a seemingly harder problem, participants became more focused on finding \textit{a solution } than finding \textit{a secure solution}. Under these circumstances, those who had access to Copilot may have benefited from a source of code (other than themselves) that placed no less (or more) a premium on secure code. While beyond the scope of this study, we discussed ways of further testing this explanation.
For RQ2, we observed a fairly uniform security performance across the different vulnerability types indicating that there was no disproportionate impact of Copilot access on the presence of any one vulnerability type.
\section{Acknowledgements}\label{sec:acknowledgements}

We would like to acknowledge and thank the reviewers for their comments and suggestions during the review process. We would also like to thank Aniruddhan Murali for his role in helping analyze participants' submissions for security vulnerabilities. This work was supported in part by the David R. Cheriton Chair in Software Systems and by the WHJIL.

\bibliographystyle{ACM-Reference-Format}
\bibliography{references}

\appendix

\end{document}